\include{Commands}
\documentclass[11pt]{article}
\usepackage{arxiv}
\usepackage{geometry}                
\usepackage{graphicx,bm}
\usepackage{epstopdf, epsfig}
\usepackage{xcolor,color}
\usepackage{amsmath}
\usepackage{mathtools}
\usepackage{amssymb}
\usepackage{amsthm}

\newcounter{defcounter}
\usepackage{hyperref}
\usepackage{cleveref}
\usepackage{comment}
\usepackage[color=green!20]{todonotes}

\usepackage[backend=bibtex,firstinits=true,style=numeric-comp,date=year,sorting=none]{biblatex}
\renewbibmacro{in:}{}
\AtEveryBibitem{\clearlist{language}}
\title{Self-organization of active rod suspensions on fluid membranes and thin viscous films}
\author{Arijit Mahapatra,
Ronit Freeman,
and Ehssan Nazockdast$^*$ \\
Department of Applied Physical Sciences, University of North Carolina at Chapel Hill,U.S.A.\\
Email addresses of corresponding authors: 
\href{ehssan@email.unc.edu}{\color{blue}{ehssan@email.unc.edu}}
}

\begin{document}

\maketitle

\begin{abstract}

Many biological processes involve transport and organization of inclusions in thin fluid interfaces. A key aspect of these assemblies is the active dissipative stresses applied from the inclusions to the fluid interface, resulting in long-range active interfacial flows. We study the effect of these active flows on the self-organization of rod-like inclusions in the interface. Specifically, we consider a dilute suspension of Brownian rods of length $L$, embedded in a thin fluid interface of 2D viscosity $\eta_m$ and surrounded on both sides with 3D fluid domains of viscosity $\eta_f$. The momentum transfer from the interfacial flows to the surrounding fluids occurs over length $\ell_0=\eta_m/\eta_f$, known as Saffman-Delbr\"uck length. We use zeroth, first and second moments of Smoluchowski equation to obtain the conservation equations for concentration, polar order and nematic order fields, and use linear stability analysis and continuum simulations to study the dynamic variations of these fields as a function of $L/\ell_0$, the ratio of active  to thermal stresses, and the dimensionless self-propulsion velocity of the embedded particles. We find that at sufficiently large activities, the suspensions of active extensile stress (pusher) with no directed motion undergo a finite wavelength nematic ordering, with the length of the ordered domains decreasing with increasing $L/\ell_0$. The ordering transition is hindered with further increases in $L/\ell_0$. In contrast, the suspensions with active contractile stress (puller) remain uniform with variations of activity. We notice that the self-propulsion velocity results in significant concentration fluctuations and changes in the size of the order domains that depend on $L/\ell_0$.  Our research highlights the role of hydrodynamic interactions in the self-organization of active inclusions on biological interfaces.

\end{abstract}

\section{Introduction}
There are many examples in biology, where the constituents are embedded and move through a thin layer that behaves as a fluid in its tangential direction. Perhaps the most important example of this is the dynamics and assembly of proteins in the cell membrane \cite{nicolson2014fluid}, which controls many cellular functions. 
In many cases membrane-protein interactions are active, i.e., they continuously perform work on the surrounding membrane.  
Coarse-grain simulations of active proteins show that 
these activities generate active force dipoles and long-range flows in the membrane \cite{huang2013coarse}. These hydrodynamic interactions (HIs) can lead to enhanced diffusion and directed motion of embedded proteins \cite{mikhailov2015hydrodynamic, hosaka2017lateral}. 
Other examples of active particles embedded in fluid interfaces include the dynamic organization of actin filaments in the cell cortex \cite{reymann2016cortical}, the growth and spread of bacteria along substrate interfaces \cite{coelho2020propagation} and the collective cell migration in tissues \cite{alert2020physical}. Many in-vitro experiments also use thin films for studying self-assembly \cite{Sanchez2012}, which effectively constrains the motion to a 2D surface.  

It is a well-known that fluid-mediated interactions are qualitatively different 
in two-dimensional (2D) and three-dimensional (3D) geometries \cite{manikantan2020surfactant}. A clear example of this is the decay of the fluid velocity field produced by a point-force in the fluid domain. In a 2D planar surface the velocity decays very slowly as $\ln (r)$ which diverges as $r\to \infty$ (known as Stokes's paradox), while it decays as $1/r$ in 3D; here, $r$ is the distance from the point-force. 
In a pioneering study Saffman and Delbr\"uck \cite{saffman1975brownian} and Saffman \cite{saffman1976brownian} studied the mobility of a disk-like inclusion in a thin film. They resolved the Stokes's paradox by assuming the membrane (or the thin fluid film) is surrounded by 3D fluid domains on one or both sides. The traction exerted by the 3D flows to the membrane introduces a length-scale, referred to as Saffman-Delbr\"uck length and defined as: $\ell_0 =\eta_m/(\eta^+ +\eta^-)$, where $\eta_m$ is the 2D membrane viscosity, and $\eta^+$ and $\eta^-$ are the 3D fluid viscosities on the top and bottom of the membrane. 
Through this addition, they were able to resolve Stokes's paradox and show that the translational mobility of a nanoscopic disk-like protein of radius $R$ in a planar membrane is only a weak logarithmic function of its radius, when $\ell_0/R\gg 1$: 
$M_\text{disk} \sim \ln(\ell_0/R)/\eta_m$; in comparison, the mobility of a sphere in 3D fluids is inversely proportional to its radius, $M_\text{sph} \sim R^{-1}$.
When $r<\ell_0$, the membrane flows induced by particle transport are dominated by membrane viscosity,  and when $r>\ell_0$ they are dominated  by flows in the surrounding fluids. 
Similarly, the scaling of particle mobility with its size qualitatively changes as the particle size becomes comparable and larger than $\ell_0$ \cite{stone2000philip, manikantan2020surfactant}. In case of rod-like particles, 
Levine et al. \cite{levine2004dynamics, levine2004mobility} showed (through simulation) that when $L/\ell_0 \ll 1$ the drag coefficients of a rod-like inclusion of length $L$ are equal in parallel and perpendicular direction and asymptote to Saffman's results for a disk: $\xi_{\parallel,\perp} \sim \eta_m \ln^{-1} (\ell_0/L)$. In contrast, when $L/\ell_0\gg 1$ the drag force is dominated by 3D fluid viscosity and the drag coefficients in parallel and perpendicular directions scale as $\xi_\parallel\sim \eta_f L/\ln(L/\ell_0)$, $\xi_\perp \sim \eta_f L$, where $\eta_f$ is the viscosity of the surrounding 3D fluid.   

Despite the clear differences between 2D vs 3D  HIs and flows, there are few studies on the consequences of these differences on the collective behavior of proteins and biopolymers on fluid biological interfaces. In particular, Manikantan tuned the effective hydrodynamic length induced by couplings with the 3D varying the depth of the adjacent 3D fluid domains, and studied its effect on the clustering of a suspension of active disks \cite{manikantan2020tunable}. Simulation results showed that increasing confinement leads to stronger clustering. Another study by the same group explored the effect of membrane non-Newtonian rheology on the clustering behavior of active disks that move under a net force, where viscosity is a function of pressure. Their simulation results show that increasing the pressure-thickening of the membrane enhances the clustering \cite{vig2023hydrodynamic}. Recently, the same group extended their studies to suspensions of rod-like suspensions under a net external force using kinetic theory \cite{manikantan2024stability}. They showed that the concentration of the suspension is unstable with perturbations, when the behavior is dominated by membrane viscous dissipation, and that increasing the dissipation from 3D fluid domain suppresses this instability. 

The effect of mechanical coupling between the active fluid film and its surrounding environment
has also been studied in simulations of active nematics by addicting a friction term that scale with fluid velocity in the momentum equation of the active film: $-f \mathbf{u}$, where $f$ is the phenomenological friction coefficient \cite{thampi2014active, doostmohammadi2016stabilization, srivastava2016negative, ardavseva2025beyond}. 
The mentioned active nematic theories assume a dense suspension of rods beyond nematic order transition and account for 
steric interactions through phenomenological free energy expressions based on symmetry arguments. 
Simulations using these models find that increasing friction leads to a decrease in the size of the ordered domains, and thus an increase in the number of observed defects, and at very high frictions the nematic ordering is suppressed. 
There are also few experimental studies of active rod-like suspensions using  \emph{in-vitro} reconstituted system of microtubules and kinesin motors. The friction in experiments is modulated by changing the  viscosity of the surrounding 
fluid \cite{guillamat2017taming} and controlled polymerization of the surrounding fluid. The experimental findings are in general agreement with the simulation results.  

This work extends the previous models of 
active rod-like suspensions on 2D fluid interfaces, surrounded by fluid domains in two important aspects. 
(i) Previous simulation of active and force-free suspensions model the membrane-surrounding fluid interactions through a  friction term: $-f \mathbf{u}$. 
In reality the traction from the surrounding fluid domains is nonlocal and more complex function 
of the surrounding fluids velocity field \cite{stone1998hydrodynamics}. The local friction model is only accurate 
when the surrounding fluids have a finite depth that
is much smaller than Saffman-Delbr\"uck length \cite{stone1998hydrodynamics,evans1988translational, sackmann1996supported}. 
In this study, instead of using a local friction model,  we solve the coupled membrane and the adjacent fluids momentum equations. This allows computing the traction force from the detailed flow field in surrounding fluids without ambiguity.
(ii) We use moment expansions of Smoluchowski microstructural theory to describe the variations of concentration, polar order and nematic order in space and time \cite{saintillan2008instabilities, saintillan2014active, saintillan2015theory}. As a result, active particles shape and size are directly included in the equations, through their translational and rotational diffusion coefficients.
This is particularly important since the functional 
forms of the drag coefficients, and thus diffusion coefficients, change with $L/\ell_0$. As a result, in addition to modulating the viscosity of the surrounding fluid, the length of the active rods can also change the size of ordered  domains and number of defects.  

We assume the membrane is incompressible and is surrounded by Newtonian fluids of the same viscosity on both sides. The active stresses are modeled using traceless contractile (puller) and extensile (pusher) force-dipoles. We consider dilute suspensions and neglect steric interactions, to focus on the coupling between membrane and 3D fluids flows. We explore the spatial variations of concentration polar and nematic orders and membrane flows as a function of $L/\ell_0$, the ratio of activity to thermal stresses and the dimensionless self-propulsion velocity of the active rods. Both linear stability analysis and  continuum simulations of the governing equations show that at sufficiently large activities the pusher suspensions with no directed motion undergo a finite wavelength nematic ordering, with the length of the ordered domains decreasing with increasing $L/\ell_0$. The ordering transition is stopped with further increase in $L/\ell_0$. 

The limit of $L/\ell_0 \to 0$ corresponds to purely 2D active rod suspension with no coupling to 3D domains. Our results in this limit show that the size of the ordered domains scale with system  scale i.e. the size of the simulation box, which is in agreement with simulations of active rod suspensions in 2D periodic geometry \cite{saintillan2008instabilities}.
The opposite limit of $L/\ell_0 \to \infty$ corresponds to 2D active rod suspensions at the interface between two fluids or fluid and air. 
Previous simulation studies in this limit also predict a finite wavelength instability \cite{Gao2015}. Our formulation and results connect these two limits through including membrane viscosity and the coupling between membrane and 3D fluid hydrodynamics. In comparison, the puller suspensions 
remain stable and uniformly distributed. We also perform simulations of self-propelled active rods. We observe that, in addition to nematic ordering, the system undergoes large concentration fluctuations. These observations are in general agreement with the previous continuum simulations of self-propelled active rods in purely 2D periodic geometries \cite{saintillan2008instabilities}. 


We note a few important assumptions made in this study to reduce the complexity of the system and focus on the effect of flow coupling between membrane and the surrounding 3D fluids. One key assumption is the planar geometry of the membrane. We know that the cell membrane is generally curved and spherical.
Several studies have shown the qualitative changes in flows and the dynamics of embedded disk-like \cite{henle2010hydrodynamics, samanta2021vortex, sigurdsson2016hydrodynamic, rower2022surface, shi2022hydrodynamics, shi2024drag} and elongated \cite{shi2022hydrodynamics, shi2024drag} proteins, when membrane is curved and enclosed.
Another key assumption is that the membrane 
does not deform and 
remains static normal to its surface. We also know that protein-membrane interactions typically coincide with membrane deformations and remodeling \cite{mcmahon2005membrane, zimmerberg2006proteins, argudo2016continuum, kozlov2014mechanisms, alimohamadi2018modeling,mahapatra2021mechanics}. 
There has been significant advancements in developing and performing continuum simulations of the 
active biological assemblies on dynamic curved surfaces \cite{al2021active, torres2019modelling, tozzi2019out, le2021dynamic, mahapatra2021, mahapatra2023formation, metselaar2019topology,mahapatrainterplay}. 
The majority of these simulations primarily focus on the role of protein-protein interactions 
in inducing membrane deformations, with less focus on the membrane flows.  
Furthermore, the number of parameters  in the simulations can be quite large and the predicted responses are even more complex. This makes isolating the role of hydrodynamics on the collective behavior very difficult. Our goal here is to study the system in its simplest form and develop the ground basis for future studies that include static and dynamic curved geometries. This will be discussed further in the \textit{Conclusion} section. 
\section{Formulation}%
\begin{figure}[!htb]
    \centering
    \includegraphics[width=0.75\textwidth]{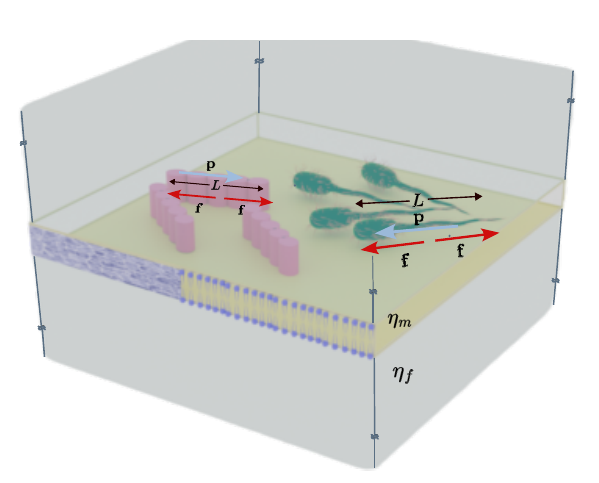}
    \caption{A schematic of a pusher (extensile) suspension of rod-like proteins and microswimmers of length $L$ on a two-dimensional lipid bilayer or a viscous film of 2D viscosity $\eta_m$ and surrounded by two semi-infinite 3D Newtonian fluid domains of viscosity $\eta_f$ on both sides. }
    \label{fig:schematic}
\end{figure}
Consider a suspension of active rod-like particles embedded in a 2D Newtonian fluid interface of 2D viscosity $\eta_m$ and surrounded on both sides by a 3D Newtonian fluid of viscosity $\eta_f$ (\cref{fig:schematic}). 
For a large assembly of particles, the position, $\mathbf{x}$, and orientation, $\mathbf{p}$, of the rod-like particles can be represented by 
their probability distribution function $\psi (\mathbf{x},\mathbf{p})$, which satisfies the following Smoluchowski conservation equation \cite{Saintillan2018anrev,Saintillan2013goveq},
\begin{equation}
\label{eq:Smoluchowski}
\frac{\partial \psi}{\partial t}+\nabla_s \cdot(\dot{\mathbf{x}} \psi)+\nabla_\mathbf{p} \cdot(\dot{\mathbf{p}} \psi)=0.
\end{equation}
Here, $\nabla_s =(\mathbf{I}-\mathbf{p}\mathbf{p})\cdot \nabla$ and $\nabla_\mathbf{p}=(\mathbf{I}-\mathbf{p}\mathbf{p})\cdot \left(\partial /\partial \mathbf{p}\right)$, and $\dot{\mathbf{x}}$ and $\dot{\mathbf{p}}$ are the translational and rotational velocities of the particle phase. For dilute Brownian rods with swimming velocity ${U}_0 \mathbf{p}$, these velocities are expressed as
\cite{saintillan2008instabilities,Saintillan2013goveq}
\begin{subequations}
\begin{align}
\label{eq:trans_vel}
\dot{\mathbf{x}}&=U_0 \mathbf{p}+\mathbf{u}-\mathbf{D}\cdot \nabla_s(\ln \psi),   \\
 \dot{\mathbf{p}}&=(\mathbf{I}-\mathbf{pp})\cdot [(\gamma \mathbf{E}+\mathbf{W})]-D_r\nabla_s(\ln \psi),
 \label{eq:roation_jeff}
\end{align}
\end{subequations}
where $\mathbf{u}$ is the 2D velocity field on the interface (membrane); $\mathbf{D}$ is the translational diffusion expressed as $\mathbf{D}=D_\parallel \mathbf{pp}+D_\perp (\mathbf{I}-\mathbf{pp})$, where $D_\parallel$ and $D_\perp$
are the rod's diffusion coefficients in parallel and perpendicular directions of its axis and $\mathbf{I}$ is the identity tensor; and $D_r$ is the  rotational diffusion coefficient of a single rod. 
The first term on the right-hand-side of Eq.~\ref{eq:roation_jeff} is the well-known Jeffrey's angular rotation, where
$\mathbf{E}$ and the antisymmetric $\mathbf{W}$ are the rate-of-strain and vorticity tensors \cite{batchelor1970slender} and $\gamma$ is the shape factor; for slender rods with large aspect ratios $\gamma \approx{1}$.  
The momentum and mass conservation for the fluid interface are
\begin{equation}
\label{eq:2d_stokes}
\eta_m \nabla_s^2 \mathbf{u}-\nabla_s q  -\nabla_s \cdot \boldsymbol{\Sigma}_p+\left(\boldsymbol{\Sigma}_f \cdot \hat{\mathbf{\nu}}\right)\vert_{z=0}=\mathbf{0},\quad \nabla_s\cdot\mathbf{u}=\mathbf{0},
\end{equation}
where $\eta_m$ is the 2D viscosity of the membrane, $q$ is the surface pressure, and $\Sigma_p$ is the stress induced by the presence of rods on the interface; the last term on the LHS of Eq.~\ref{eq:2d_stokes} models the traction applied from 3D flows to the interface, where $\boldsymbol{\Sigma}_f$ is the 3D fluid stress and $\hat{\mathbf{\nu}}$ is the unit normal vector of the planar interface. The 3D fluid flows are described by the Stokes and continuity equations:
\begin{equation}
\eta_f \nabla^2 \mathbf{u}_f-\nabla p= \mathbf{0},\quad \nabla \cdot \mathbf{u}_f=0. 
\label{eq:3d_stokes}
\end{equation}
where $\eta_f$ and  $\mathbf{u}_f$ and $p$ are the 3D fluid viscosity, velocity and pressure fields. 
The boundary conditions (BCs) are zero 3D fluid velocity far from the interface ($\mathbf{u}_f =\mathbf{0}\vert_{z \to \pm \infty}$) and no-slip  
BC at the interface: $\left(\mathbf{u}_f=\mathbf{u}\right)\vert_{z=0}$. 

The velocity field due to a point force on the membrane, $\mathbf{f} (\mathbf{x}_0)$, 
can be calculated by the known Green's function of the coupled Eqs.~\ref{eq:2d_stokes} and \ref{eq:3d_stokes} that also satisfies the mentioned BCs:
$\mathbf{u}_m (\mathbf{x})=\mathbf{G}(\mathbf{x}-\mathbf{x}_0)\cdot \mathbf{f} (\mathbf{x}_0)$ \cite{levine2004dynamics}. Similarly, the membrane velocity field induced by active stresses can be expressed as
\begin{equation}
    \mathbf{u}_m(\mathbf{x})=\int \mathbf{G}(\mathbf{x}-\mathbf{x}_0)\cdot \left[\nabla \cdot \boldsymbol{\Sigma}_p (\mathbf{x}_0)\right]d\mathbf{x}_0. 
    \label{eq:integral}
\end{equation}

In the dilute limit, the stress induced by excluded volume interactions is negligible, and 
the particle stress can be approximated by only the  active stress. The active stress induced by each rod, $\mathbf{S}$, is the symmetric moment of force distribution along the rod's axis, and can expressed generally as $\mathbf{S}=\sigma \mathbf{p p}$ \cite{saintillan2014active}. The sign of the force-dipole $\sigma$ denotes if the stress is contractile along the axis  or a puller swimmer ($\sigma>0$) or extensile (pusher, $\sigma<0$). The traceless part of the volume average active stress can be computed as
\begin{equation}
\boldsymbol{\Sigma}_p( \mathbf{x} , t)=\sigma \int_S \psi( \mathbf{x} , \mathbf{p} , t)\left( \mathbf{p p} -\frac{\mathbf{I}}{2}\right) d \mathbf{p}. 
\label{eq:activestress}
\end{equation} 
 
The system of equations \ref{eq:Smoluchowski}-\ref{eq:activestress} can be solved numerically to determine 
$\psi$, $\mathbf{u}$, $q$, $\mathbf{u}_f$ and $p$. For 2D geometries, $\psi$ is a function of orientation $\theta$, and position on the interface ($x$ and $y$). Solving PDEs with even three independent variables can be numerically expensive. A useful method for improving numerical efficiency and physical interpretation is to use zeroth, first and second moments of $\psi$ with respect to $\mathbf{p}$, 
corresponding to concentration field, $c(\mathbf{x},t)$, polar order parameter, $\mathbf{n}(\mathbf{x},t)$ and nematic order parameter, $\mathbf{Q}(\mathbf{x},t)$ \cite{saintillan2014active}: 
\begin{equation}
\begin{aligned}
c( \mathbf{x} , t) & =\int_S \psi( \mathbf{x} , \mathbf{p} , t) d \mathbf{p}, \\
 \mathbf{n} ( \mathbf{x} , t) & =\frac{1}{c( \mathbf{x} , t)} \int \mathbf{p} \psi( \mathbf{x} , \mathbf{p} , t) d \mathbf{p}, \\
\mathbf{Q} ( \mathbf{x} , t) & =\frac{1}{c( \mathbf{x} , t)} \int\left( \mathbf{p p} -\frac{\mathbf{I}}{2}\right) \psi( \mathbf{x} , \mathbf{p} , t) d \mathbf{p}.
\end{aligned}
\label{eq:moments}
\end{equation}
%
%
%
The governing equations for $(c,\mathbf{n},\mathbf{Q})$ are obtained by applying expression in Eq.~\ref{eq:moments} to Eq.~\ref{eq:Smoluchowski}. 
\begin{equation}
\label{eq:gov_nd}
\begin{aligned}
D_t \phi= & -\hat{{U}}\nabla \cdot(\phi \mathbf{n})+\frac{1}{Pe_\perp}\nabla \cdot \left(\left[\frac{D_\parallel}{D_\perp} \mathbf{n}\mathbf{n}+ (\mathbf{I}-\mathbf{nn})\right]\cdot  \nabla \phi \right) \\
D_t(\phi \mathbf{n})= & -\hat{{U}}[\nabla \cdot(\phi \mathbf{Q})+\frac{1}{2} \nabla \phi] \\
& +\frac{1}{Pe_\perp} \nabla \cdot \left(\left[\frac{D_\parallel}{D_\perp} \mathbf{n}\mathbf{n}+ (\mathbf{I}-\mathbf{nn})\right]\cdot  \nabla (\phi \mathbf{n})\right) \\
&+(\phi \mathbf{I} \mathbf{n}-\langle\mathbf{p p p}\rangle):(\beta \mathbf{E}+\mathbf{W})- \frac{2}{Pe_r} \phi \mathbf{n} \\
D_t(\phi \mathbf{Q})= & -\hat{{U}}\left[\nabla \cdot\langle\mathbf{p p p}\rangle-\frac{\mathbf{I} }{ 2} \nabla \cdot(\phi \mathbf{n})\right] \\
& +\frac{1}{Pe_\perp} \nabla \cdot \left(\left[\frac{D_\parallel}{D_\perp} \mathbf{n}\mathbf{n}+ (\mathbf{I}-\mathbf{nn})\right]\cdot  \nabla (\phi \mathbf{Q})\right) \\
&+\beta \phi[\mathbf{E} \cdot(\mathbf{Q}+\mathbf{I} / 2)+(\mathbf{Q}+\mathbf{I} / 2) \cdot \mathbf{E}]  +\phi[\mathbf{W} \cdot \mathbf{Q}-\mathbf{Q} \cdot \mathbf{W}] \\
&-2 \beta\langle\mathbf{p p p p}\rangle: \mathbf{E}- \frac{4}{Pe_r} \phi \mathbf{Q},
\end{aligned}
\end{equation}
where $D_t = \partial/\partial t +\mathbf{u} \cdot \nabla $ denotes material derivative.  
In deriving the above equation We have used the following scaling relations to non-dimensionalize the governing equations:
\begin{align*}
    & \mathbf{x} \sim L,& & c \sim \phi/L^2, & & \Sigma_p \sim 
 |\sigma|/L^2 ,& & {t \sim \tau=\eta_m L^2/|\sigma|,} &
 & {\mathbf{u} \sim L/\tau}&
\end{align*}
with $L$ defining the rod's length, $Pe_\perp=|\sigma|/\eta_m D_\perp$, and $P_r=|\sigma|/D_r \eta_m L^2$ are the respective Peclet numbers in lateral and rotational directions, and $\hat{U}$ is the dimensionless self-propulsion velocity. 
The dimensionless form of Eqs.~\ref{eq:2d_stokes} and \ref{eq:fluid_3D} are 
\begin{subequations}
  \begin{align}
\label{eq:fluid_mem}
    &\nabla^2_s{\mathbf{ {u}}_m} -\nabla_s {q} -\nabla_s\cdot (\phi \mathbf{Q})+\left(\frac{L}{\ell_0}\right)\left[ \nabla \mathbf{{u}}_f+\nabla^T\mathbf{{u}}_f\right]\cdot\hat{\mathbf{\nu}}=\mathbf{0},\\
    \label{eq:fluid_3D}
    &{\left(\frac{L}{\ell_0}\right)}\nabla^2\mathbf{{u}}_f-\nabla {p}=\mathbf{0}.  
\end{align}
\label{eq:fluids}
\end{subequations}

%
%
%
%
%
%

As expected, coarse-graining through moment expansion leads to terms involving higher order moments, i.e. $\langle \mathbf{ppp}\rangle$ and $\langle \mathbf{pppp} \rangle$. 
Closure models are typically used to approximate these higher order moment in terms of $\mathbf{Q}$ and $\mathbf{n}$.
Here, we use the closure model by Doi et. al \cite{Doi1981}, which in Einstein notation is written as
\begin{subequations}
\begin{align}
   &  \left\langle p_i p_j p_k\right\rangle \approx \frac{c}{4}\left(n_i \delta_{j k}+n_j \delta_{i k}+n_k \delta_{i j}\right),\\
    &  \left\langle p_i p_j p_k p_l\right\rangle \approx \left( c Q_{{i,j}} + \frac{I_{{i,j}}}{2}  \right) \left( c Q_{{k,l}}  + \frac{I_{{k,l}}}{2} \right),
\end{align}
\end{subequations}
where $\ell_0=\eta_m/\eta_f$ is the well-known Saffman-Delbr\"uck 
length. 

The thermal diffusion coefficients are determined by fluctuation-dissipation theorem, $D_{(\parallel, \perp, R)}=k_B T \xi^{-1}_{(\parallel, \perp, r)}$, where $\xi$ is the hydrodynamic drag coefficients of a single rod in parallel, perpendicular and rotational directions.  
In 3D Stokes flow, the drag coefficients are given by
$\xi^\text{3D}_\perp \approx 2 \xi_\parallel^\text{3D} \approx 4\pi\eta_f /\ln (L/a)$, where $L/a \gg 1$ is the  rod's aspect ratio, and so $D_\parallel^\text{3D}/D_\perp^\text{3D} \approx 2$. 
In comparison, the drag of a rod moving in a viscous film or a lipid membrane \cite{levine2002dynamics} can be expressed as $\xi^\text{2D}_\parallel/2\pi \eta_m=\mathcal{F}_\parallel(L/\ell_0)$, 
$\xi^\text{2D}_\perp/4\pi \eta_m=\mathcal{F}_\perp(L/\ell_0)$, and $\xi_r^\text{2D}/4\pi \eta_m L^2=\mathcal{F}_r(L/\ell_0)$, 
where $\mathcal{F}_{(\parallel, \perp, r)}$ are different functions of $L/\ell_0$ that can be numerically calculated.
In the limit of $L/\ell_0 \ll 1$ the drag is dominated by membrane viscous forces and these functions asymptote to  $\mathcal{F}_{(\parallel, \perp)} \approx \ln^{-1}(\ell_0/L-0.5772)$ and $\mathcal{F}_r\approx 1.48/2\pi$. In the opposite limit of $L/\ell_0$ the drag is dominated by the traction applied from the 3D flows and we have
$\mathcal{F}_\parallel \approx  0.25 (L/\ell_0)\ln^{-1}(0.43 L/\ell_0)$, $\mathcal{F}_\perp \approx  0.25 (L/\ell_0)$ and $\mathcal{F}_r \approx  (1/16\pi)(L/\ell_0)$. So we have 
$D_\parallel/D_\perp \approx 1$ when $L/\ell_0 \ll 1$ and $D_\parallel/D_\perp \approx \ln (0.43 L/\ell_0)$ when $L/\ell_0 \gg 1$. This means that unlike 3D flows $D_\parallel/D_\perp$ change with the length of the filament. 
We fit power function in terms of $L/\ell_0$ to the computed values of $\mathcal{F}_{(\parallel, \perp, r)}$ over the range of $L/\ell_0 \in [10^{-3}-10^{3}]$. These expressions are provided in the SI. 

The governing equations \ref{eq:gov_nd}, are characterized by four dimensionless variables: $\phi$, $\hat{\sigma}=|\sigma|/k_B T$, $\hat{L}=L/\ell_0$, and $\hat{U}$. In the next section we investigate the influence of these variables through linear stability analysis and numerical simulations. 
\section{Linear Stability analysis}
\label{sec:linstability}
\begin{figure*}[!htb]
    \centering
    \includegraphics[width=0.9\textwidth]{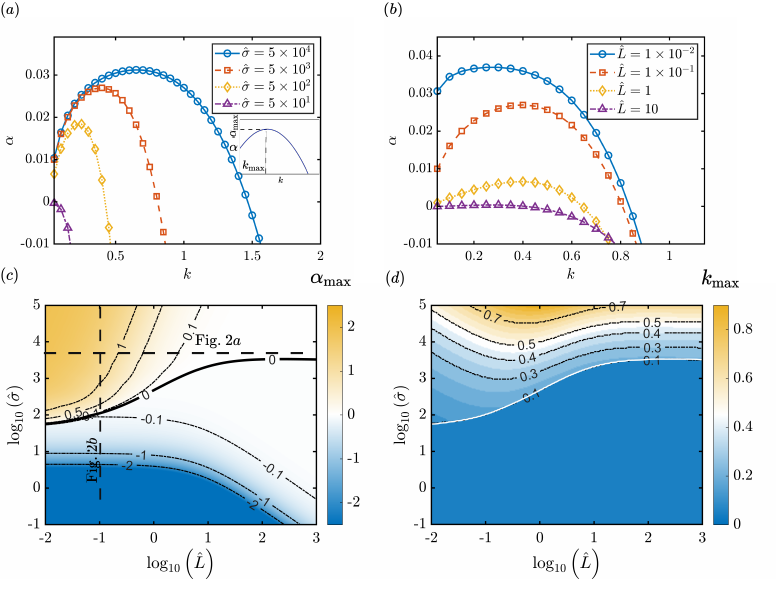}
    \caption{\textbf{Linear stability analysis of non-propelling pusher suspensions: the effect of the dimensionless activity and length.} (a-b) Variations of the growth rate (eigenvalue) with wave number at (a) different activity ratios and $\hat{L}=0.1$, and (b) at $\hat{\sigma}=5\times 10^3$ and different $\hat{L}$. (c-d) Phase diagram of (c) the growth rate and (d) the wave number corresponding to maximum unstable mode, as a function of dimensionless length and activity.}
    \label{fig:stability_swim}
\end{figure*}
We perform a linear stability analysis of the transport equations (\cref{eq:gov_nd}) to find the parameter space where hydrodynamic flows drive ordering in concentration, polar and nematic fields.  We consider perturbations around a homogeneous distribution of rods $(\phi=\phi_0)$ with no preferred orientation $(\mathbf{n}=\mathbf{0}, \mathbf{Q}=\mathbf{0})$:
\begin{equation}
    \phi=\phi_0\left(1+\epsilon \phi^{\prime}\right), \quad \mathbf{n}= \epsilon \mathbf{n}^{\prime}, \quad \mathbf{Q}=\epsilon \mathbf{Q}^{\prime},\quad \mathbf{u}_{(m,f)}=\epsilon\mathbf{u}^\prime_{(m,f)},
\end{equation}
where $\epsilon \ll 1$. The equations for $(\phi^\prime, \mathbf{n}^\prime, \mathbf{Q}^\prime)$ reduce to 
\begin{subequations}
\begin{align}
\label{eq:lin_den}
\frac{\partial \phi^{\prime}}{\partial t} =&  - \hat{{U}}\left(\nabla \cdot \mathbf{n} ^{\prime}\right)+Pe^{-1}_\perp \nabla^2 \phi^{\prime}, \\
\label{eq:lin_n}
\frac{\partial \mathbf{n}^{\prime}}{\partial t}=&  -\hat{{U}}\left[\left(\nabla \cdot\mathbf{Q}^{\prime}\right)+\frac{1}{2}\nabla \phi^{\prime}\right]+Pe^{-1}_\perp \nabla^2( \mathbf{n}^{\prime})-Pe^{-1}_r \mathbf{n} ^{\prime},\\
\label{eq:lin_Q}
\frac{\partial \mathbf{Q}^\prime}{\partial t}= & -\frac{\hat{{U}}}{5}\left[\nabla \mathbf{n}^{\prime} + (\nabla \mathbf{n}^{\prime})^T-\frac{3 \mathbf{I}}{2} \left(\nabla \cdot \mathbf{n}^{\prime}\right)\right] \\
&\nonumber + Pe_\perp^{-1} \nabla^2 \mathbf{Q}^{\prime} +\frac{\beta }{2}\mathbf{E}^{\prime}  -4 Pe_r^{-1} \mathbf{Q}^{\prime}.
\end{align}
\end{subequations}
The form linearized equations reveal some key features of the system's dynamics.  
The equations, and by extension the system's stability, are independent of 
$\phi_0$ to the first order of $\epsilon$. Furthermore the advection terms in the material derivative, i.e., $\mathbf{u}\cdot \nabla (\phi,\mathbf{n},\mathbf{Q})$, are of $\mathcal{O}(\epsilon^2)$ and, thus, do not appear in the linearlized equations.  The fluid flows still contribute through $\mathbf{E}^\prime$ term in the equation for $\mathbf{Q}\prime$. 
Finally, the system of equation contains only perpendicular Peclet number, $Pe_\perp$, showing the system's stability is independent of parallel drag up to $\mathcal{O}(\epsilon)$. These features are also observed in active rod suspensions in 3D \cite{saintillan2014active}.

We use a periodic computational domain to model the infinite 2D planar geometry. Thus, it is more convenient to write down the equations in Fourier space:
\begin{equation}
    \left(\phi^\prime, \mathbf{n}^\prime, \mathbf{Q}^\prime\right)=\sum_k\left(\phi_k, \mathbf{n}_k, \mathbf{Q}_k\right) e^{\alpha_k t+i 2\pi \mathbf{k}\cdot \mathbf{x}}.
\end{equation}
Upon substitution, Eqs. \ref{eq:lin_den}-\ref{eq:lin_Q}) simplify to
\begin{equation}
\begin{aligned}
\label{eq:gov_k}
    \alpha_k \phi_k =&   -2 \pi \hat{{U}} i  \mathbf{k} \cdot \mathbf{n} -4 \pi^2 k^2 Pe^{-1}_\perp \phi_k, \\
     \alpha_k \mathbf{n}_k=& -2 \pi \hat{{U}} i \left(\mathbf{k} \cdot \mathbf{Q}_k+\frac{1}{2} \mathbf{k} \phi_k\right)-4 \pi^2 k^2 Pe^{-1}_\perp \mathbf{n}_k- Pe^{-1}_r  \mathbf{n}_k, \\
   \alpha_k \mathbf{Q}_k =& -\frac{2 \pi \hat{{U}} i }{5}\left( \mathbf{k} \mathbf{n}_k +  \mathbf{n}_k \mathbf{k}-\frac{3}{2} \mathbf{I} \mathbf{k} \cdot \mathbf{n}_k\right)\\
&-4 \pi k^2  Pe^{-1}_\perp \mathbf{Q}_k +\frac{\beta }{2}\mathbf{E}_k   -4 Pe^{-1}_r  \mathbf{Q}_k.
\end{aligned}
\end{equation}

The membrane velocity field is calculated by taking a Fourier transform of eq.~\ref{eq:integral} and using convolution theorem:
\begin{align}
    &(\mathbf{u}^\prime_m)_k=\mathbf{G}_k \cdot \left(2\pi \mathbf{k} \cdot \mathbf{Q}^\prime\right),& 
    &\mathbf{G}_k=\frac{1}{4 \pi^2 ( k^2\ell_0/L+k)}\left(\mathbf{I}-\frac{\mathbf{k k}}{k^2}\right).&
\end{align}
Substituting for $\mathbf{G}_k$ in the above equation we get  
\begin{equation}
\label{eqn:Hasimoto}
    (\mathbf{u_m}^\prime)_k=\frac{1}{2 \pi( k^2 \ell_0/L+k)}\left( \mathbf{k} \cdot \mathbf{Q}_k-\frac{\left(\mathbf{kk} : \mathbf{Q}_k\right) \mathbf{k}}{k^2}\right).
\end{equation}
This expression is used to calculate $\mathbf{E}_k$. 
The linear system of equations contain three dimensionless parameters: $\hat{L}$ and $\hat{\sigma}$, and $\hat{U}$ ($Pe_\perp$ is a function of $\hat{L}$ and $\hat{\sigma}$). 
\Cref{eq:gov_k}, can be written as an eigenvalue equation:
\begin{equation}
[\Lambda -\alpha \mathbf{I}]\left[\begin{array}{l}
\phi_k \\
\mathbf{n}_k \\
\mathbf{Q}_k
\end{array}\right]=\mathbf{0}.
\end{equation}
We compute the eigenvalues of $\Lambda$ in terms of $\hat{L}$ and $\hat{\sigma}$ and $\hat{U}$.
Non-negative eigenvalues are regions of instability. 
\subsection{Pullers:}
We find that, similar to dilute purely 3D and 2D active puller rods \cite{saintillan2008instabilities, saintillan2015theory}, puller suspensions are unconditionally stable. Moving forward, we will focus on the behavior of pusher (extensible) rods.
We begin by studying the case of no self-propulsion i.e. shakers, which is more applicable to assemblies of cytoskeletal filaments and proteins, including actin filaments, microtubules and septin proteins, 
on the cell membrane. Later in section \S \ref{sec:Simulation} we study the effect of swimming velocity on the collective behavior, which more closely represents bacterial swimmers in thin films. 
\subsection{Pushers with no self-propulsion,  $\hat{U}=0$:}

\Cref{fig:stability_swim}(a) shows the computed eigenvalue, $\alpha$, vs wavenumber $k$ for diffident values of $\hat{\sigma}\in [50-5\times10^3]$, when $\hat{L} =0.1$.  At low activities ($\hat{\sigma}= 50$) $\alpha$ becomes negative irrespective of $k$, predicting that the perturbations decay over time to the uniform distribution. 
The eigenvalues become positive with increasing activity and follow a non-monotonic variations with $k$; the maximum 
growth rate ($\alpha_\text{max}$) occurs in a finite wavenumber ($k_\text{max}$) that increases with activity, suggesting a dominant size for the unstable phase that decreases with activity.
This behavior is qualitatively different from purely 2D and 3D active suspensions, where the maximum growth is observed at $k=0$, corresponding to instabilities that scale with the system size \cite{saintillan2008instabilities}.
\begin{figure}[!tbh]
    \centering
    \includegraphics[width=0.95\textwidth]{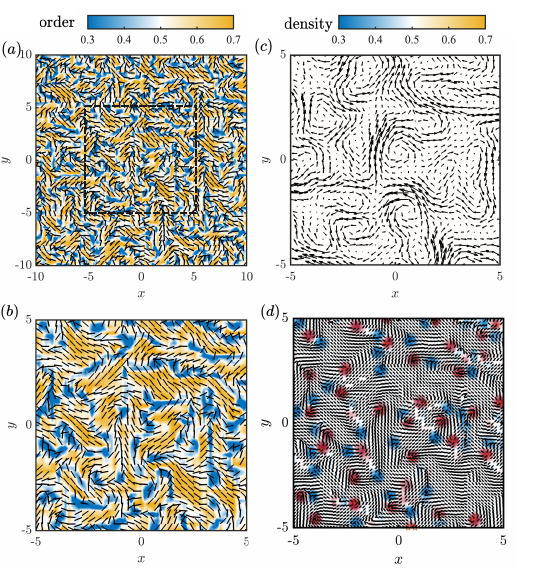}
    \caption{\textbf{Snapshots of the continuum simulations of the director, fluid velocity and concentration fields of non-propelling pusher suspensions in the planar periodic domain at $\hat{U}=0$, $\hat{\sigma}=5\times 10^{3}$ and $\hat{L}=0.1$, and all taken at the same simulation time.} (a)  Distribution of director field with scalar order parameter shown as the background heatmap, with $1$ corresponding to fully ordered and $0$ corresponding to fully disordered. (b) Enlarged view of the selected box in subplot(a). (c) The distribution of the velocity field at the interface with the concentration distribution shown as the background heatmap. The uniform white color background denotes uniform concentration. (d) Defect points corresponding to the same snapshot, red triangle are the $+1/2$ defects and blue circle are $-1/2$ defects. }
    \label{fig:no_swim_mor}
\end{figure}
\Cref{fig:stability_swim}(b) shows the non-monotonic variations of $\alpha$ vs $k$ for different values of $\hat{L}$ and $\hat{\sigma}= 5\times 10^{3}$. We see that $\alpha$ is decreased with increasing $\hat{L}$
and ultimately becomes negative at sufficiently large values of dimensionless length; see the curve for $\hat{L}=10$ in \Cref{fig:stability_swim}(b). To explain this trend, it is useful to explore the decay velocity field  around a point-force in the membrane (or a viscous film).  When $\hat{L} \ll 1$, the fluid flows are dictated by membrane viscosity and the velocity field decays as $\ln(r)$ in all directions, as is the case in 2D stokes flow. When $\hat{L} \ge 1$ the velocity field in the direction of  the applied force decays as $1/r$, and it decays as $1/r^2$ in the perpendicular
direction\cite{manikantan2020tunable, levine2004dynamics}. 
By extension, the velocity field induced by a stresslet decays as $1/r$ when $\hat{L} \ll 1$, and as $1/r^2$ and $1/r^3$ in parallel and perpendicular directions, when $\hat{L} \gg 1$. 
We see that as $\hat{L}$ is increased the hydrodynamic interactions are weakened and become more short-ranged. Thus, larger activities are needed at larger values of $\hat{L}$ (stronger hydrodynamic screening) to drive the hydrodynamic instability. 
\begin{figure*}[!hbt]
    \centering
    \includegraphics[width=\textwidth]{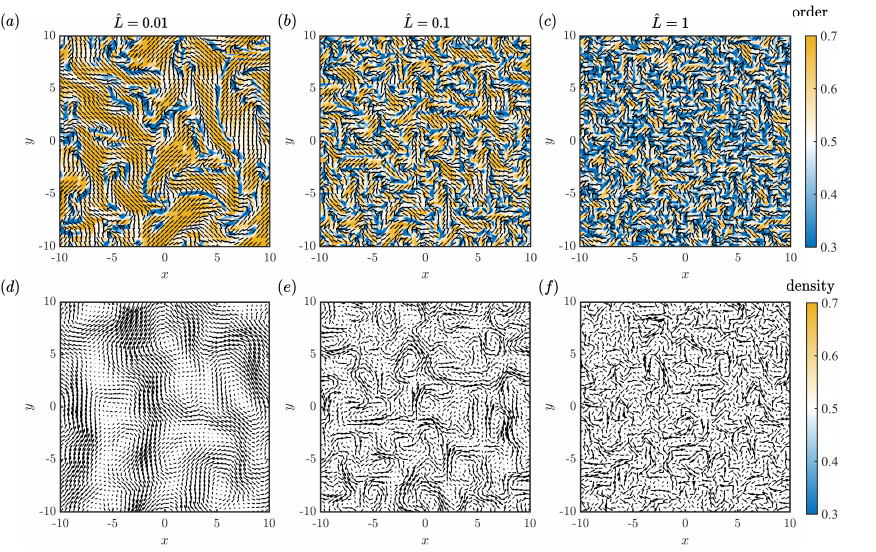}
    \caption{\textbf{Snapshots of continuum simulations of non-propelling pusher suspensions at $\hat{U}=0$, $\hat{\sigma}=5\times 10^{3}$ and $\hat{L}=0.01,\, 0.1$ and $1$.} The top row shows the distribution of the director field and scalar order parameter (background heatmap) and the bottom row shows the fluid velocity and concentration (background heatmap) fields corresponding to the same $\hat{L}$. The size of the ordered domains and coherent fluid flows decrease from $\hat{L}=0.01$ to $\hat{L}=0.1$. At $\hat{L}=1$ the director field appears almost random and and the fluid flows appear turbulent in the scale of the rod's length (1 in simulations). }
    \label{fig:no_swim_Ll0}
\end{figure*}
Furthermore, we see $k_\text{max}$ is increased as the dimensionless length is increased from $\hat{L} = 10^{-2}$ to $\hat{L} =0.1$, and remains roughly the same for $\hat{L}=1$. In the limit of $\hat{L} \to 0$, the size of instabilities scale with system size ($k_\text{max} \to 0$). 
This can be explained by noting that when $\hat{L} =0$, the membrane fluid flows are entirely governed by membrane viscosity the there is no physical length scale in the problem other than system size. In this limit  the equations reduce to dilute 2D active nematic suspensions for which $k_\text{max} =0$ \cite{saintillan2014active}.

\Cref{fig:stability_swim}(c) shows a contour plot of the maximum growth rate as a function of the dimensionless length and activity. The solid line shows the marginal stability ($\alpha=0$), that separates the stable and unstable
regions in the parameter space.
Note that the contours are closer when $\hat{L}< 1$, and they widen for larger values of the length ratio, depicting a more significant energy barrier for instability for higher length ratios.
This behavior can be explained by noting that the perpendicular drag 

The threshold activity number at marginal stability increases with a higher length ratio; this increase happens due to momentum transfer to bulk fluid and would not appear in the system without explicit dependence of $\hat{L}$. 
The phase diagram (2c) shows that the marginal stability line ($\alpha_{max}$) reaches a plateau both at the high and low length ratio, where the transport is dominated by membrane viscosity and bulk viscosity, respectively. 
In the dispersion relation (2a,b), one can see that the maximum value of growth-rate alpha corresponds to some intermediate value of k ($=k_{max}$).
In \Cref{fig:stability_swim}(d), we presented a contour plot of the maximum unstable wavenumber with dimensionless activity and length ratio. 
We observe $k_{max}$ varies non-monotonically with length ratio; however, with activity, it increases. 
\section{Continuum Simulations}
\label{sec:Simulation}
a In this section, we present the continuum simulations of active rod suspensions described by the system of equations in \ref{eq:gov_nd}, coupled to the membrane flow field described by \Cref{eq:fluid_mem}. 
We use a finite difference and implicit-explicit time-stepping \cite{Mahapatra2020} to discretize Eqs. \ref{eq:gov_nd}, and Fourier spectral methods \cite{camley2013diffusion} for computing $\mathbf{u}_m(\mathbf{x})$.
The velocity field in Fourier space is given by
\begin{equation}
\mathbf{u}_k=\frac{1}{4 \pi^2\left(\ell_0 k^2+k\right)}\left(\mathbf{f}_k-\frac{\left(\mathbf{k} \cdot \mathbf{f}_k\right) \mathbf{k}}{k^2}\right),
\label{eq:spectral}
\end{equation}
where $\mathbf{f}_k= i 2 \pi \mathbf{k} \cdot \mathbf{Q}_k$, is the force field in the Fourier space, and is treated explicitly in time. $\mathbf{u}_m(\mathbf{x})$ is computed by computing $\mathbf{Q}_k$ from the previous time through Fourier transform, using \Cref{eq:spectral} to compute $\mathbf{u}_k$, and finally taking the inverse Fourier transform to get $\mathbf{u}_m(\mathbf{x})$.
We use python 3 for numerical simulation in $64 \times 64$ grid points with
time step of $\Delta t= 5 \times 10^{-3}$.

\subsection{No self-propulsion:} 

We begin with presenting the results for the case of no self-propulsion ($\hat{{U}}=0$). In this case, the 
net polar order is zero in average, and the local alignment
is described by the nematic order parameter, $\mathbf{Q}$.
Specifically, the degree of the local alignment can be measured by scalar 
order parameter, which in 2D systems is simply the largest eignenvalue of $\mathbf{Q}$ tensor. The rod's local orientation is given by the eigenvector corresponding to this eignevalue.  
{\Cref{fig:no_swim_mor}}a shows a snapshot of local orientation of continuum simulations in the periodic domain, for the choice of $\hat{\sigma} = 5 \times 10^3$ and $\hat{L} = 0.1$, after the system has undergone ordering transition. The background heatmap shows the scalar order parameter, where regions with a higher scalar order parameter (highlighted in yellow) correspond to areas of highly ordered director fields, indicating parallel orientations of the rods. The vector field show the normalized orientation of the rods. Note that the vectors have no specific direction (no arrows), since the rods have no inherent polarity in non-swimming suspensions. A zoomed-in plot of the dashed square region is shown in {\Cref{fig:no_swim_mor}}b. We observe that the self-assembly occurs at a finite length scale, consistent with the maximum unstable wave number for this parameter set (Fig.\ref{fig:stability_swim}). 

{\Cref{fig:no_swim_mor}}c shows a snapshot of the membrane velocity field induced by active stresses, corresponding to the orientation fields shown in {\Cref{fig:no_swim_mor}}b. 
The velocity field forms circulation patches of the same size as the patches in the space order parameter and the orientation field. 
The rod concentration field is also shown in the same figure as a heatmap in the background. As it can be seen, the concentration field remains uniform throughout the domain ($\phi=0.50$, corresponding to the background white color). 

{\Cref{fig:no_swim_mor}}d displays defects and discontinuities in the director field. The defects are characterized by the winding number, $W$, calculated using the following equation [needref]. 
\begin{equation}
\label{eq:winding}
W=\frac{1}{2 \pi} \oint_{ C (\kappa)}\left[\hat{n}_1(\kappa) \frac{\partial \hat{n}_2(\kappa)}{\partial \kappa}-\hat{n}_2(\kappa) \frac{\partial \hat{n}_1(\kappa)}{\partial \kappa}\right] d \kappa.
\end{equation}
Here $\kappa$ is the coordinate along a closed curve $\oint_{ C (\kappa)}$ enclosing the point. For our calculation, we considered ${ C (\kappa)}$ as a circle of radius twice the grid size centered around the finite difference grid points. $\hat{n}_1$ and $\hat{n}_2$ are the components of the director field on the curve. We numerically performed the integration with 2D grid points on the curve ${ C (\kappa)}$. 
Red patches in {\Cref{fig:no_swim_mor}}d represent $+1/2$ defects, and blue patches represent $-1/2$ defects. 
These defect pairs are predominantly found between regions of self-assembled rods.
%
%
\begin{figure}[!htb]
    \centering
    \includegraphics[width=0.95\textwidth]{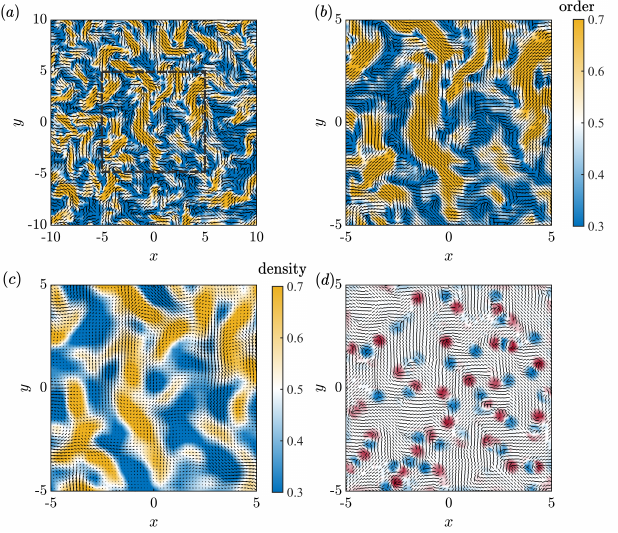}
    \caption{\textbf{Self-propulsion leads to concentration fluctuations}. (a) Snapshots of continuum simulations of (a) the director field and scalar order parameter (background heatmap); (b) the zoomed-in view of the square region in subplot (a). (c) The distribution of interface velocity and concentration (background heatmap) fields corresponding to subplot (b).   (d) The distributions of defects corresponding to subplot (b). The results were obtained using $\hat{U}=1$, $\hat{L}=0.1$ and $\hat{\sigma}=5\times 10^3$.}
    \label{fig:swim_mor}
\end{figure}
\begin{figure*}[!tbh]
    \centering
    \includegraphics[width=0.95\textwidth]{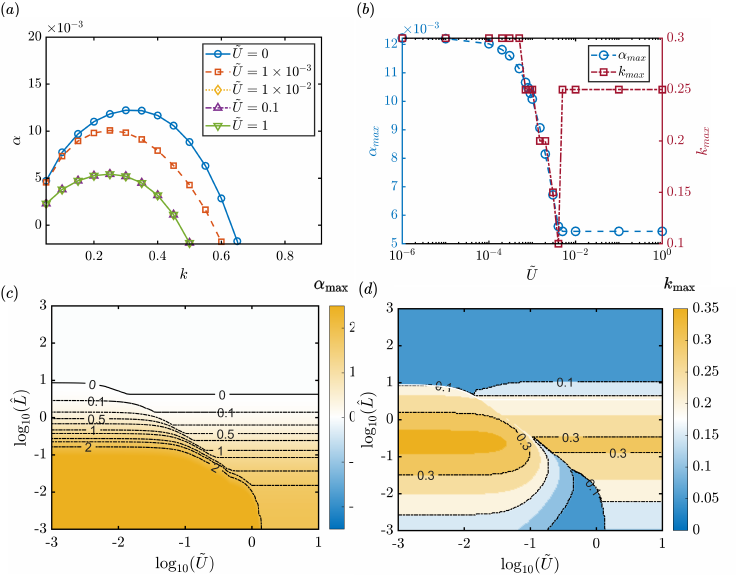}
    \caption{\textbf{Linear stability analysis: the effect of self-propulsion speed.}  (a) Growth rate vs wavenumber at $\hat{L}=0.1$, $\hat{\sigma}=5\times 10^{3}$ and self-propulsion velocities. (b) Maximum growth rate ($\alpha_\text{max}$) and the wavenumber corresponding to that growth rate vs dimensionless self-propulsion velocity for $\hat{\sigma}=2\times 10^{3}$ and  $\hat{L}=0.1$. (c-d) Heatmaps of the maximum growth rate (c) and the wave vector corresponding to it (d) as a function of 
    $\hat{U}$ and $\hat{L}$. 
    }
    \label{fig:swim_Ll0}
\end{figure*}

\begin{figure*}[!hbt]
    \centering
    \includegraphics[width=0.95\textwidth]{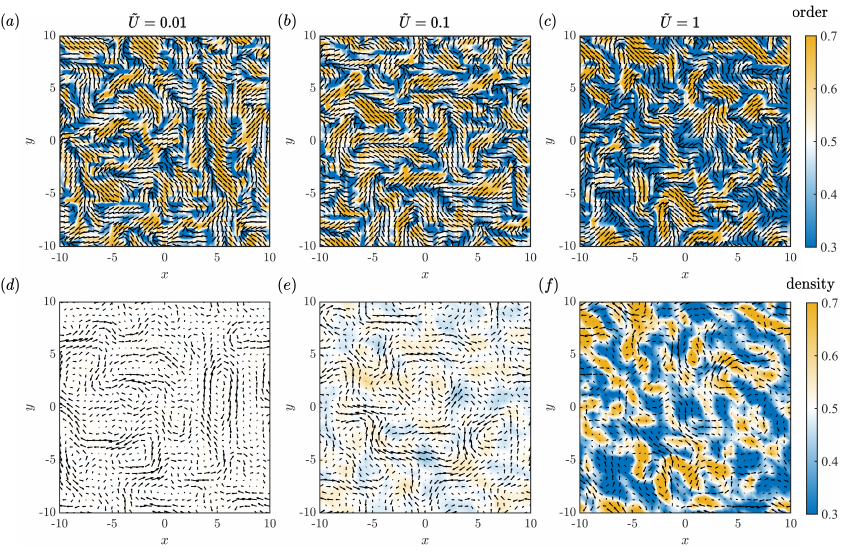}
    \caption{\textbf{Snapshots of continuum simulations of non-propelling pusher suspensions at $\hat{L}=0.1$, $\hat{\sigma}=2\times 10^{3}$ and $\tilde{U}=0.01,\, 0.1$ and $1$.} The top row shows the distribution of the director field and scalar order parameter (background heatmap) and the bottom row shows the fluid velocity and concentration (background heatmap) fields corresponding to the same $\tilde{U}$. }
    \label{fig:swim_Usm}
\end{figure*}
Next, we investigate the changes in orientation and velocity fields of active rod suspensions with varying $\hat{L}$. 
The continuum simulations (\Cref{fig:no_swim_Ll0}a-c) show that the size of the ordered domains reduces with increasing $\hat{L}$, which is in line with the results of linear stability analysis (Fig.~\ref{fig:stability_swim}). 
For lower values of $\hat{L}$ ($\hat{L}=0.01$), we observe that the size of the self-assembled structure spans over the computational domain (\Cref{fig:no_swim_Ll0}a), which demonstrates the long wavelength instability observed in active rod-like suspensions in a periodic 2D domain \cite{saintillan2008instabilities}.
For the intermediate value of the length we observe a finite-size ordered domain of the rods, where bulk and membrane viscosity both contribute to the hydrodynamics. 
In comparison when $\hat{L}=1$, ordering and coherent flows are suppressed in scales larger than the rod's length (larger than 1 in dimensionless form), as shown in  \cref{fig:no_swim_Ll0}c and \cref{fig:no_swim_Ll0}f. Again, these observations are in agreement with the linear stability analysis presented earlier (\cref{fig:stability_swim}d). 
We note that the coarse-grain kinetic model used here
is suitable for studying structural and flow features that 
are larger than the size of the particles. 
This assumption breaks down, at $\hat{L} > 1$ (\cref{fig:no_swim_Ll0}c and \cref{fig:no_swim_Ll0}f). Thus, we have limited our analysis here to $\hat{L} \le 1$. 
As we shall discuss, $\hat{L}$ is satisfied in most 
physiological conditions involving the assembly of proteins and biopolymers on the cell membrane and other biological interfaces.

\subsection{Self-propulsion:}
Next, we explore how self-propulsion velocity can alter the collective dynamics. 
Figures \ref{fig:swim_mor}(a-b) shows the predicted
scalar order parameter at $\tilde{U}=1$, $\hat{L}=0.1$ and $\hat{\sigma}=5\times 10^3$, where $\tilde{U}=4\pi \mathcal{F}_\perp \hat{\sigma} \hat{U}$. We chose $\tilde{U}$ as the dimensionless variable to study the self-propulsion because this is the combination of variables that appears in the linear stability analyses. 
The qualitative features of the spatial variations of the scalar order parameter  and the fluid velocity field are quite similar to active suspensions with no self-propulsion (\Cref{fig:swim_mor}a,b).
However, unlike the simulations with no self-propulsion, we observe large spatial fluctuations of concentration in length scales larger than the rod's length in self-propelled systems (\Cref{fig:swim_mor}c). 
This is consistent with the earlier simulations of 2D dilute self-propelled rods without any coupling to 3D fluid domains that also show concentration fluctuations \cite{saintillan2008instabilities}. 

A more subtle difference with simulations with no self-propulsion (Fig. 4(b) and 4(e)) is the reduction in the number of domains (corresponding to reduction in $k_\text{max}$) in simulations of self-propelled suspensions. To explore this in more details, we studied the change in the linear stability of the system with variations of $\tilde{U}$; the results are shown in Fig.~\ref{fig:swim_Ll0}.  
As shown in Fig.~\ref{fig:swim_Ll0}, for $\hat{\sigma}=5\times 10^3$ and $\hat{L}=0.1$, the maximum growth rate decreases with increasing $\tilde{U}$ and occurs at lower $k_\text{max}$. This is consistent with the continuum simulation results in Fig.~\ref{fig:swim_mor}. 
We also observe that the $\alpha$ vs $k$ 
converge to two distinct asymptotic limits at 
$\tilde{U} \ll 1$ and $\tilde{U} \gg 1$, corresponding to diffusion-dominant and advection-dominant translational motion, respectively. 
\par
Fig.~\ref{fig:swim_Ll0} shows the variations of $\alpha_\text{max}$ and $k_\text{max}$ with changes in $\tilde{U}$, for $\hat{L}=0.1$ and $\hat{\sigma} =2\times10^3$. Again, we see that both $\alpha_\text{max}$ and $k_\text{max}$ reduce with increasing $\tilde{U}$ and asymptote to well-defined limits at $\tilde{U} \ll 1$ and $\tilde{U} \gg 1$. 
Interestingly, we also observe a sharp decrease in $k_\text{max}$ followed by an even sharper increase, in a small range of $\tilde{U}\sim \mathcal{O}(1)$. 
Figs.~\ref{fig:swim_Ll0}(c-d) extend the results of Figs.~\ref{fig:swim_Ll0}(a-b) to different values of $\hat{L}$ and show the heatmaps of $\alpha_\text{max}$
and $k_\text{max}$ as a function of $\tilde{U}$ and $\hat{L}$. We see that the findings 
in Figs.~\ref{fig:swim_Ll0}(a-b) generally hold for a wide range of $\hat{L}$. In particular, we can clearly see the local minima in $k_\text{max}$ with variations of $\tilde{U}$, when $\hat{L} \le 1$, which corresponds to the blue and white regions of the heatmap in Fig.~\ref{fig:swim_Ll0}(d).  We note that the eigenvalues corresponding to $\alpha_\text{max}$ have no imaginary component within numerical error (imaginary parts are smaller than $10^{-15}$). 

Furthermore, we performed continuum simulations in the parameter space that includes the computed local minimum in $k_\text{max}$ in
linear stability analysis, taking $\hat{\sigma}=2\times 10^{2}$ and $\hat{L}=0.1$ and varying $\tilde{U}$. The
results are shown in Fig.~\ref{fig:swim_Usm}. We did not find any signatures in the number of domains in simulation results that correspond to the computed local minimum in $k_\text{max}$ in linear stability analysis. 
As it can be seen, when $\tilde{U}\ll 1$ we observe nearly uniform concentration distribution and the size and number of ordered domains remain unchanged with variations of $\hat{U}$. At $\tilde{U}=1$ we observe aggregation domains that closely (but not exactly) correspond to  the ordered domains. Moreover, we observe that at $\hat{U}=1$ the ordered and disordered domains become sharper and more distinct, compared to simulations at lower self-propulsion velocities. 

\begin{figure*}[!htb]
    \centering
    \includegraphics[width=0.9\textwidth]{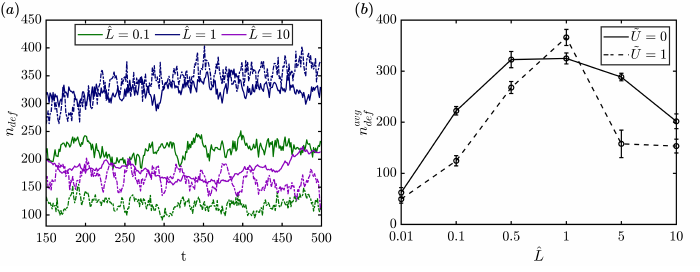}
    \caption{\textbf{Effect of self-propulsion velocity on defect number and dynamics.} (a) Time variations of the number of defect points in pusher suspensions with no-self-propulsion (rigid lines) and those with $\tilde{U}=1$ (dashed lines) at different values of $\hat{L}$. (b) Average steady-state number of defects vs $\hat{L}$ for non-propelling and self-propelled ($\hat{U}=1$) pusher suspensions. The number of defects initially increase and then decrease with $\hat{L}$. Aside from $\hat{L}=1$, the number of defects is smaller (domains are larger) in self-propelled suspensions, which is in agreement with the results of linear stability analysis. }
    \label{fig:def_all}
\end{figure*}

To get a more quantitative understanding of the variations ordered domains with self-propulsion velocity, we computed
the changes in the number of defects with time at different $\hat{L}$ in systems with no self-propulsion and $\tilde{U}=1$ all at $\hat{\sigma}=5\times 10^3$; the results shown in \Cref{fig:def_all}. As shown in \Cref{fig:def_all}(a), the number of defects reaches a dynamic steady state at $t>350$. Regardless of self-propulsion velocity, we see a non-monotonic variations of the number of defects with $\hat{L}$: we see an increase in the 
number of defects as $\hat{L}$ is increased from $0.1$ to $1$, and then the number of defects shows a large drop at $\hat{L}=10$. Tis trend is more clearly observed in \Cref{fig:def_all}(b), which shows the average number of defects vs $\hat{L}$
for $\tilde{U}=0$ and $1$. Note that aside from $\hat{L}=1$, the number of defects is consistently smaller in self-propelled 
systems. This is consistent with the predictions of $k_\text{max}$ from linear stability analysis in \Cref{fig:swim_Ll0}(b) and \ref{fig:swim_Ll0}(d). 

\section{Discussion and Conclusions}
We presented a continuum model for simulating the self-organization of suspensions of active rod-like particles that are embedded in a thin viscous interface and surrounded by semi-infinite 3D fluid domains on both sides. 
Following prior studies, the active stress was modeled using a simple force-dipole in the rod's main axis ($\mathbf{S}=\sigma \mathbf{p}\mathbf{p}$, where $\mathbf{p}$ is the rod's main axis), with $\sigma<0$ denoting a puller rod and $\sigma>0$ denoting a pusher rod.  
For dilute suspensions with negligible steric interactions, the collective behavior is determined by 
three dimensionless parameters: the dimensionless activity, $\hat{L}$ and the dimensionless self-propulsion velocity.
The conservation equations for the concentration, polar order and nematic order fields were obtained by taking the zeroth, first and second moment expansions of Smoluchowski 
conservation equation for the probability density of states. Linear stability analysis of the governing equations showed that at sufficiently high activities, pusher suspensions with no self-propulsion undergo a nematic ordering transition with a finite wavelength. These results also showed that the 
size of the ordered domains is reduced with increasing $\hat{L}$ and that further increase in $\hat{L}$ hindered 
the ordering transition and resulted in a uniform ordering 
in the scale of the rod's length. 
In the absence of self-propulsion the concentration field 
remained uniform over the entire range of parameters studied for pusher and puller active suspensions. 
In comparison, at sufficiently large self-propulsion velocities and activities the pusher suspensions undergoes large concentration fluctuations, in addition to nematic ordering. 
Linear stability analysis and continuum simulations show that the critical activity required for  ordering transition increases and the size of the ordered domains decreases with increasing self-propulsion speed.  

Our work highlights the importance of hydrodynamic coupling between interfacial flows and the surrounding environment
on the collective dynamics of active rod-like suspensions and in particular in setting the length scale of the ordered domains and concentration fluctuations. 
Our results are in qualitative agreements with simulations 
of active nematics with substrate friction \cite{thampi2014active, doostmohammadi2016stabilization, srivastava2016negative, ardavseva2025beyond}. One benefit of the current theory is that the 
conservation equations were obtained directly from microstructural Smoluchowski theory.  As a result, particle-scale parameters such as the rod's length and diffusivity are directly included in the final equations. Another feature of our model is that the coupling with the surrounding fluids was computed by solving the coupled momentum equations for the interface and the fluids. 
This is particularly important, since
the rod's translational and rotational mobility, and by extension diffusivity, are distinct functions of $\hat{L}$. 
At $\hat{L} \ll 1$, the fluid dissipation is dominated by interface (membrane) viscosity and the translational mobility, and thus diffusivity, is only a weak logarithmic function of length. In contrast, when $\hat{L} \ge1$ the dissipation is dominated by the surrounding 3D fluids and mobility scales inversely with length. 
Thus, the size of the ordered domains can also change with changing the rod's length. 

In many biological interfaces, including the cell membrane,  the interface is curved and enclosed,  which adds the dimension of the enclosed geometry, say $R$, as another hydrodynamic length scale.  
The measurements of the cell membrane viscosity 
vary greatly in the range $\eta_{m}\in[10^{-10}, 10^{-6}]\,(\mathrm{Pa}\cdot\mathrm{s}\cdot\mathrm{m})$, depending on the measurement technique and lipid  composition \cite{block2018brownian, sakuma2020viscosity, nagao2021relationship}. Taking the surrounding fluid to be water, we get $\ell_0 \in [10^{-6},\,10^{-3}]$ (m). Assuming cell dimensions to be $\mathcal{O}(\mu m)$, we may have $R< \ell_0$ in many instances. 
In these conditions the smallest hydrodynamic length that couples interfacial and the surrounding fluid flows is $R$.
In the special case of a spherical geometry, the fundamental solutions to singularities can be obtained in closed form \cite{henle2010hydrodynamics}, and so the flows generated by active stresses can be computed using eq.~\ref{eq:integral}. Furthermore, the fundamental solutions can be used in a slender-body formulation to compute the mobility of the rods as a function of $\hat{L}$ and $L/R$, where $R$ is the radius of the sphere \cite{shi2022hydrodynamics}. 
Previous simulation studies \cite{shi2022hydrodynamics} show that when the rod's length is much smaller than the sphere's radius, the rod's mobility on a spherical membrane can be  mapped to the planar membrane values, as long as $\ell_0$ is replaced by $\ell^\star=\min(\ell_0,R)$. We expect that the collective behavior of active rods would also map to the planar membrane results presented here, as long as $L/\ell^\star \ll 1$. 
The same studies also show a sharp increase in the rod's 
perpendicular drag coefficient on a spherical membrane, compared to a planar membrane, when the rod's length becomes comparable to the sphere's radius:
$L/R \sim \mathcal{O}(1)$. 
This enhanced resistance to motion 
arises from the flow confinement effects in the enclosed 
spherical geometry \cite{shi2022hydrodynamics}. 
These large differences are expected to 
change the phase diagram for nematic ordering and the size of the ordered domains.  
Note, however, that the continuum approximation becomes 
less accurate as the rod's length is increased to be comparable or larger than the sphere's radius.
Discrete particle simulations can be used as the alternative to study the collective behavior in this limit.

In some biological processes 
and physical applications the fluid interface is next to 
a solid-like substrate, such as the membrane attachment to the cell cortex, or a 3D fluid domains of a finite depth.
This change in boundary reduces the effective hydrodynamic coupling length compared to $\ell_0$. Again, the fundamental solutions for singularities in the presence of a rigid boundary have been obtained for planar \cite{stone1998hydrodynamics} and 
spherical \cite{shi2024drag} geometries, allowing the same methodology to be applied to these problems. 
In the limit of strong frictional forces between the interface and the surrounding substrate, the perpendicular drag coefficient 
scale as $\xi_\perp \sim L^2$ \cite{shi2024drag}, which is qualitatively different from the drag coefficient scaling in planar membrane.
While we expect the qualitative features of the collective dynamics to remain similar to the planar membrane, we also expect significant quantitative differences in the
parameters regime that cause ordering as well as the size of the ordered domains.

\section*{Conflicts of interest}
The authors declare no competing interests. 

\section*{Acknowledgements}
AM and EN acknowledge support by the National Science Foundation grant 1944156. AM, EN and RF acknowledge funding support by the Alfred P. Sloan Foundation grant G-2021-14197. 



 \clearpage
 \printbibliography


\end{document}